\documentclass[sigconf,c10pt]{acmart}

\AtBeginDocument{%
	}

\usepackage{array}
\usepackage{url}
\usepackage{lipsum}
\usepackage{multirow}
\usepackage{adjustbox}
\usepackage{mathtools}
\usepackage[ruled, vlined, linesnumbered]{algorithm2e}
\SetKwComment{Comment}{$\triangleright$\ }{}

\usepackage{caption}
\usepackage{subcaption}
\usepackage{balance}

\usepackage{pgfplots}
\usepackage{tikz}
\usetikzlibrary{arrows}
\usetikzlibrary{shapes}
\newcommand{\provtovec}{\texttt{Prov2vec}}
\newcommand*\encircle[1]{\tikz[baseline=(char.base)]{
		\node[shape=ellipse, draw,fill=green!45, inner sep=1pt, minimum width=0.2cm, minimum height=0.1cm](char){#1};}}

\newtheorem{definition}{Definition}

\begin{document}
%
\title{Prov2vec: Learning Provenance Graph Representation for Unsupervised APT Detection}

\author{Bibek Bhattarai}
\affiliation{%
	\institution{The George Washington University} 
	\country{Washington, DC, USA}}
\email{bhattarai\_b@gwu.edu}

\author{H. Howie Huang}
\affiliation{%
	\institution{The George Washington University}
	\country{Washington, DC, USA}}
\email{howie@gwu.edu}

\begin{abstract}
Modern cyber attackers use advanced zero-day exploits, highly targeted spear phishing, and other social engineering techniques to gain access and also use evasion techniques to maintain a prolonged presence within the victim network while working gradually towards the objective. To minimize the damage, it is necessary to detect these Advanced Persistent Threat as early in the campaign as possible. This paper proposes, {\provtovec}, a system for the continuous monitoring of enterprise host's behavior to detect attackers' activities.
It leverages the data provenance graph built using system event logs to get complete visibility into the execution state of an enterprise host and the causal relationship between system entities. It proposes a novel provenance graph kernel to obtain the canonical representation of the system behavior, which is compared against its historical behaviors and that of other hosts to detect the deviation from the normality. These representations are used in several machine learning models to evaluate their ability to capture the underlying behavior of an endpoint host. We have empirically demonstrated that the provenance graph kernel produces a much more compact representation compared to existing methods while improving prediction ability.

\end{abstract}


\maketitle

%

\section{Introduction}
Large enterprises and government networks have seen a significant rise in targeted attacks from experienced cyber criminals with substantial financial backing~\cite{yahoo2013,deputydogforum,target2013,solarwind2020}. These sophisticated attacks often referred to as advanced persistent threats, are carried out in multiple steps over prolonged period of time, each step designed to blend in with benign activity. These attacks bypass traditional signature-based defense mechanisms~\cite{signature} via the use of zero-day exploits. While anomaly detection processes~\cite{du2017deeplog} can detect the events that diverge from the norm, the temporal sequence-based context can not reliably capture the interrelations between involved entities and thus miss the `low and slow' attacks.

The endpoint detection and response (EDR) systems~\cite{carbonblack,falcon,esetprotect,fsecure,msendpoint} continuously monitoring and detecting threats have been helpful in reducing the dwell time of an attack from 101 days in 2017 to 16 days in 2022~\cite{mandiant2022}.
However, EDR systems, which prioritize recall, generate huge amount of false positives and low fidelity alerts. Analysts must evaluate these alerts, which involves a tedious job of writing long ad-hoc queries to Security Information and Event Management (SIEM) system. Understaffed security defense teams face an insurmountable alert volume, increasing the risk of critical alerts being overlooked due to alert fatigue~\cite{alertfatigue}.

\textit{Provenance Graphs} have gained significant attention in security research. By capturing the information flow between system objects Directed Acyclic Graph (DAG), these graphs provide historical context and the impact of an alert through backward and forward graph traversal~\cite{king2003backtracking}. Formally, we define a provenance graph snapshot at time $t$ as $G_t = (V_t, E_t, A^v, A^e)$, where $V_t$ and $E_t$ are set of nodes and edges at time $t$ and $A^v$ and $A^e$ are functions that maps all nodes $v \in V_t$ and edges $e \in E_t$ in the graph to set of node and edge attributes respectively. To enhance threat detection accuracy and provide better visibility into attack campaigns, several threat detection systems~\cite{milajerdi2019holmes,milajerdi2019poirot,hassan2020tactical,bhattarai2022steinerlog} combine provenance graph with alert generation capabilities of EDR systems. This integration allows for automated alert correlation, reducing false positives and providing contextual information around alerts. However, leveraging fine-grained analysis for detecting realistic `low and slow` APT campaigns poses various challenges.

First, performing fine-grained causal analysis on lengthy APT campaigns presents significant computational challenges. With a median dwell time of three weeks and the generation of several terabytes of log events daily, holding the entire provenance graph in memory is highly impractical. Storing graphs in databases and conducting graph traversal ~\cite{neo4j} for alert correlation significantly escalates the cost. Achieving real-time detection using compute-intensive alert correlation across the entire enterprise network is a daunting task. Furthermore, the majority of computational resources allocated to alert correlation contribute minimally to the detection process. In a study conducted by~\cite{bhattarai2022steinerlog}, involving a network with 500 hosts, only 28 hosts were compromised over a three-day evaluation period. Out of these, detection was only relevant for five hosts (1\% of total hosts), indicated by high-risk scores in the attacker's activity subgraphs. While alert correlation is crucial for retracing the attackers' steps and comprehending attacks on compromised hosts, applying it universally across all hosts and at all times is excessive.

Second, the detection capability of alert correlation-based systems is inherently limited by the coverage of rules employed in the alert generation process. These rules are manually crafted by security practitioners, relying on the cyber threat intelligence reports from security forums, blogs, social media, and previous attacks. The matching semantics are very different based on the platform and underlying sensors used for data collection. Replicating such a system in a new platform requires a substantial manual effort and expertise. While incorporating novelty detection on a node or edge level~\cite{bhatia2020midas,ranshous2016scalable,eswaran2018sedanspot} can potentially detect previously unseen attacks,  it is important to note that new benign activities also emerge constantly, necessitating a broader perspective on activities for decision-making.


To tackle these challenges, we introduce {\provtovec}, an innovative system that leverages a novel \textbf{provenance graph kernel} to derive a canonical form for a given graph snapshot, capturing the aggregated host behavior at a specific time point in a fixed-size vector representation. {\provtovec} operates by mining label-aware backward walks, with a maximum length specified by the user as \textbf{h}, for each node in provenance graph. 
These walks encompassing the execution history of nodes over varying hop lengths from $0 \le i \le h$  are then compressed into a label that succinctly describes the nodes causal history. A node label histogram is constructed by tallying the frequencies of distinct labels across all nodes in the graph for each hop length from $0$ to $h$. These histograms are stored in memory using a fixed-size probabilistic data structure called histosketch~\cite{yang2017histosketch}, which are utilized by downstream machine learning tasks to model the behavior of the hosts and detect when they behave abnormally.



This work makes several key contributions:
\begin{itemize}
	\item We develop an end-to-end system for unsupervised APT detection. Leveraging the provenance graphs built from logs gathered during normal operations, our system creates comprehensive host behavior profiles. Any provenance graphs deviating from those  generated by benign activities are identified as anomalies. We utilize these anomalous graph snapshots, along with their associated users and hosts, to traverse the authentication graph and uncover all compromised entities.
	
	\item We propose a novel graph kernel that enhances the generalization of similar provenance graph structures using compact node label histograms. Our approach achieves superior or comparable accuracy in downstream machine learning tasks while maintaining the histogram size an order magnitude smaller than Wesfeller-Lehman subtree (WLSubtree) kernel~\cite{shervashidze2011weisfeiler} and temporally ordered WL subtree kernel from Unicorn~\cite{han2020unicorn}.
	
	\item We showcase the effectiveness of {\provtovec} in profiling host behavior and detecting compromised network entities through three machine learning tasks -- graph classification, graph clustering, and graph anomaly detection -- on provenance graphs generated from Windows and Linux hosts.
	
\end{itemize}

The rest of the paper is organized into five major sections. Section~\ref{sec:threatmodel} intoduces the threat model for our system. Section~\ref{sec:p2vdesign} discusses the detail of {\provtovec} system design, where we discuss the novel provenance graph kernel for obtaining the compact node label histogram. We discuss the performance of {\provtovec} to model the enterprise host's behavior and compare it against state-of-the-art graph kernels in Section~\ref{sec:p2vevaluation}. The sections following discuss the assumptions made and shortcomings of the {\provtovec} system, summarizes the related works, and discuss their relation with {\provtovec} respectively.

\section{Threat Model}
\label{sec:threatmodel}
We focus on a typical APT life cycle, where adversaries gain unauthorized access to the enterprise hosts and aim to remain stealth for an extended period. To achieve their objectives, attackers carry out various post-exploitation activities, including internal reconnaissance, privilege escalation, lateral movement, and data exfiltration~\cite{mitreattack}. Our goal is to detect compromised hosts based on a given snapshot of the provenance graph at a specific time $t$, using {\provtovec}. We assume that {\provtovec} has sufficient historical data to establish a behavior profile of enterprise hosts during normal operations. We also assume that the provenance graph obtained during an attack exhibits distinct differences from the graphs observed during prior normal operations.

{\provtovec} does not make assumptions about the specific actions performed by an attacker, apart from the fact that  their intent and/or actions leave indicators in the audit logs and, consequently on the provenance graph. To accurately capture this information, {\provtovec} assumes the correctness of log collection frameworks. The remainder of this paper assumes the validity of log collecting frameworks and log data  used in our experiments, focusing on {\provtovec}'s ability to model system behavior based on them. For modeling the system behavior, this work assumes that provenance graphs with similar structures indicate comparable operational behavior. Therefore, the detection of abnormal behavior entails the computation  of (dis)similarities among the provenance graph snapshots.

\section{{\provtovec} Design}
\label{sec:p2vdesign}

\begin{figure}[t]
	\begin{center}
		\includegraphics[width=0.45\textwidth ]{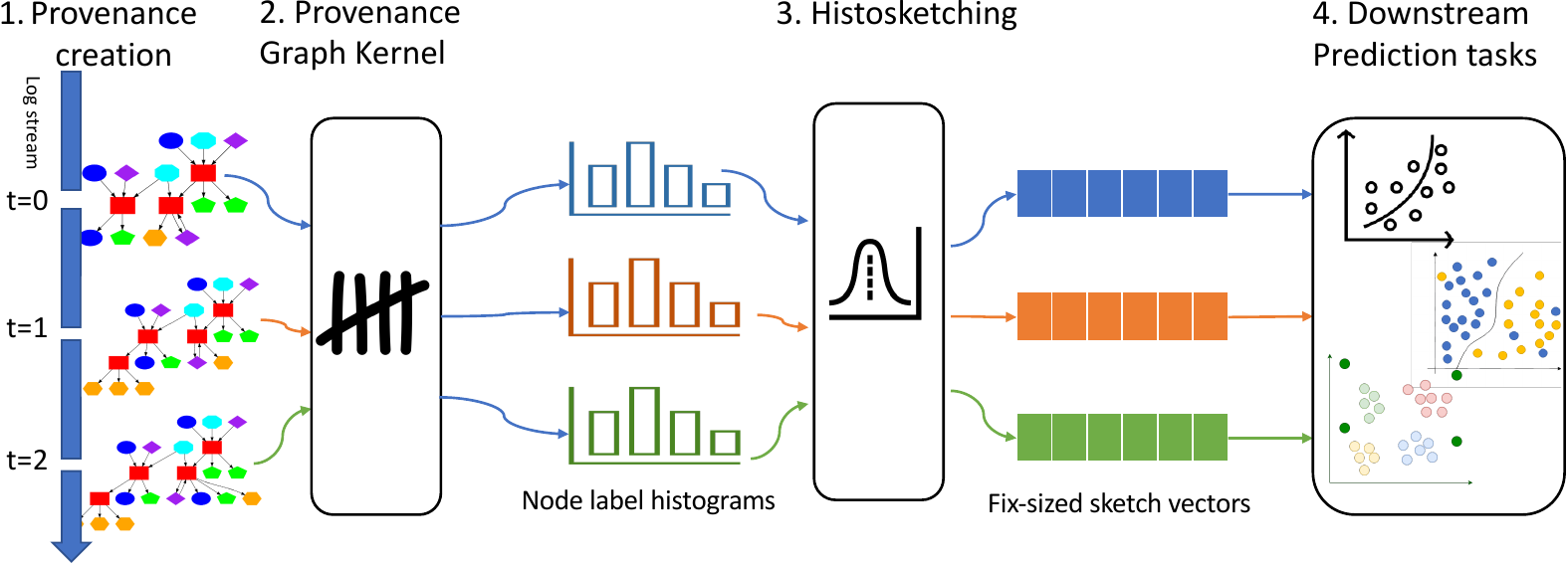}
		\caption{The system diagram of {\provtovec}.}
		\label{fig:provsystem}
	\end{center}
\end{figure}

Figure ~\ref{fig:provsystem} shows the high level overview of {\provtovec} system. \encircle{1} Given the stream of log events generated by auditing tools~\cite{auditd,dtrace,etw}, {\provtovec} updates the provenance graph continuously with new events. Periodically, it takes snapshots of said provenance graph $G_t = (V_t, E_t)$. \encircle{2} The novel provenance graph kernel is used to convert the graph snapshots into node label histograms. These histograms can have different sizes depending on the number of distinct provenance labels in a given graph snapshot while aggregating over the specified neighborhood size. \encircle{3} To compare the histograms with one another, we convert them into vectors of the same size. In the static setting, this can be done by building a vector of size equal to node label vocabulary built using the histograms of all graphs in question. In the streaming setting, the vocabulary size is constantly increasing. To enable an easy comparison of streaming histograms, we utilize a probabilistic data structure called histosketch~\cite{yang2017histosketch} that uses the consistent weighted hashing~\cite{li20150} to sample the histograms into a fixed size vectors while preserving the similarity between them. \encircle{4} Finally, the series of feature vectors representing provenance graph snapshots is fed to machine learning models to learn the behavior of an enterprise host. They can be designed for one of many tasks such as graph classification, outlier detection, and graph clustering. During deployment, the first three steps are performed and the resultant feature vector is tested against the model learned to detect whether the behavior at any instance is anomalous.

The anomalies generated from these models serve as the leads for analysts, providing indications of potentially malicious activity. These anomalies prompt further investigation to gain insights into the underlying causes and potential countermeasures. To extract subgraphs that capture the sequence of actions performed by the attacker, alert generation and correlation can be conducted using systems like Rapsheet~\cite{hassan2020tactical} or SteinerLog~\cite{bhattarai2022steinerlog}.  {\provtovec} plays a crucial role in identifying suspicious endpoint hosts, enabling alert correlation systems to focus their fine-grained analysis on those specific hosts. The remainder of this chapter delves into the various steps involved in constructing an end-to-end system for detecting compromised enterprise hosts using {\provtovec}, providing a comprehensive understanding of each component.



\subsection{Provenance Graph Creation}
\begin{figure}[t]
 	\begin{center}
 		\includegraphics[width=0.45\textwidth ]{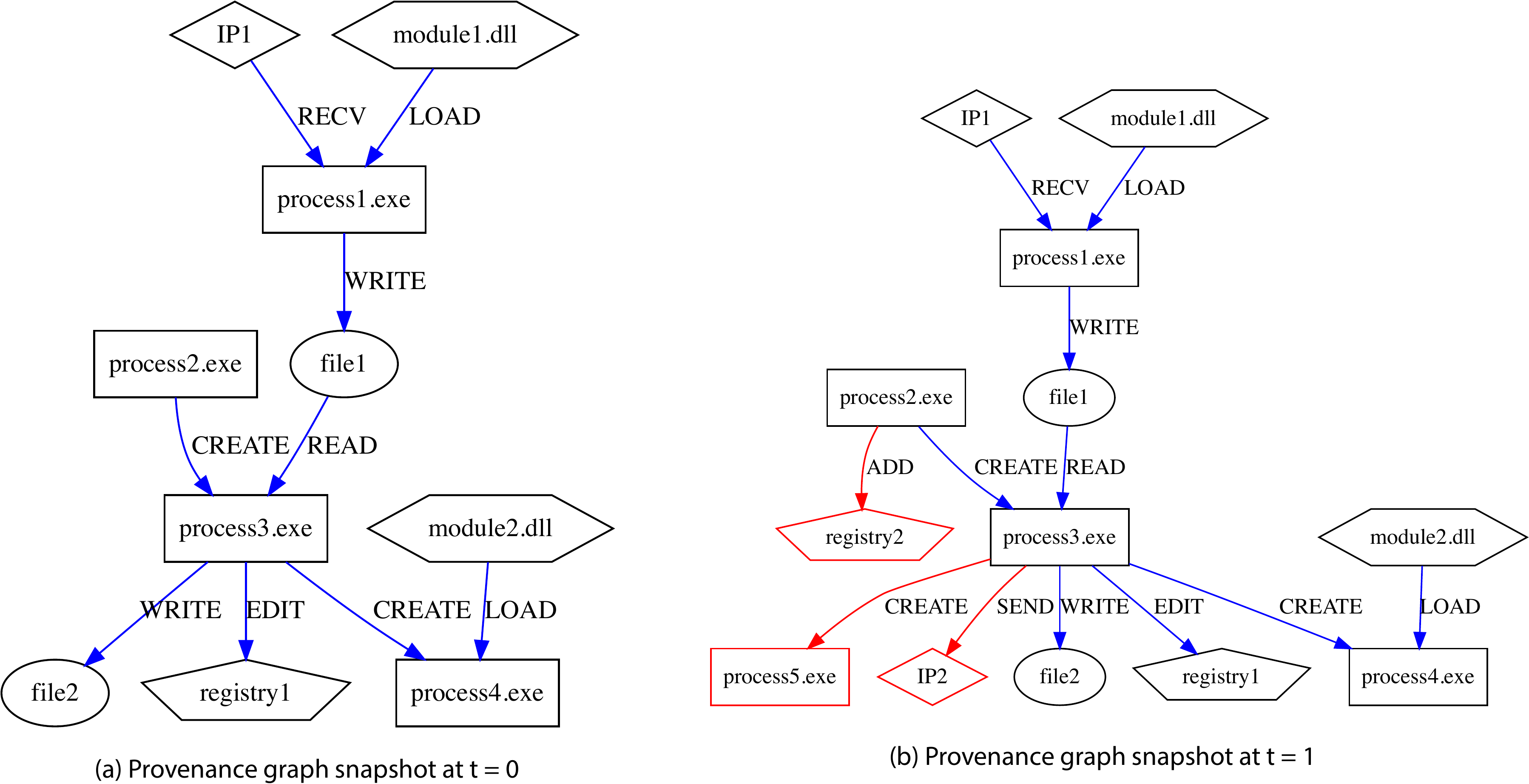}
 		\caption{The sample provenance graph captured. The graph on the right(b) has some new nodes and edges added since the last snapshot at left (a) denoted by red color. }
 		\label{fig:provsample}
 	\end{center}
 \end{figure}

The system logs are parsed into a triplet of $(subject, action, object)$ and inserted into a provenance graph. The direction of edges signifies the flow of data or information. For instance, an edge corresponding to a process writing on a file will point from the process to the file, whereas a process reading a file will have the opposite direction. Figure~\ref{fig:provsample} shows two snapshots of a provenance graph at time $t = 0$ and later at $t = 1$. The red edges and nodes on the second snapshot represent the part inserted after the first snapshot.

To reduce the graph size and avoid dependency explosion during the forensic analysis, we utilize causality preserving duplicate elimination~\cite{xu2016high} and node versioning. When inserting an edge $(src, event, dst)$, if there already is another edge with the same triplet in the provenance graph, and there are not any outgoing edges from the latest version of $dst$, i.e., $dst_i$, we simply update the time information of edge and avoid inserting the edge again. However, if the latest version of the node $dst$, i.e., $dst_i$ already has outgoing edges, the insertion of the edge changes the provenance of all those nodes. In that case, we simply create a new version of that node $dst_{i+1}$, and insert the edge  $(src, event, dst_{i+1})$ instead. In addition, an edge needs to be inserted between $dst_i$ and $dst_{i+1}$ to indicate that the latter is the newer version of the former node. In our experiment data, an average of 1.2 node versions are created for subjects and objects but we were able to reduce the number of edges in the graph by a factor of 3.38$\times$.

Node versioning and redundant edge elimination allow for efficient incremental computation of node-label histograms. By creating different versions of nodes in the provenance graph as subjects or objects change, we can focus the computation only on newly inserted nodes. This approach minimizes redundant processing and improves computational efficiency, ensuring the node-label histograms are efficiently updated as the provenance graph evolves.


\subsection{Provenance Graph Kernel}
To capture the heterogeneity of the provenance graph, we perform label-aware backward walks from each node in the given snapshot. These walks traverse the graph up to a user-defined length \textbf{h}. By accumulating the provenance labels of all nodes, we construct a provenance label histogram for the snapshot. This approach allows us to capture the diverse characteristics of the graph and generate a comprehensive representation of the node labels within the specified walk length.

\begin{definition}
	\textbf{Label-aware backward walk:} Given a node $v \in V_t$, a backward walk of length $i$ starting at $v$ is defined as \{$l(e_0)$, $l(e_1)$, $l(e_2)$, ... $l(e_{i-1})$, $l(u)$\}, where ($e_{i-1}$, $e_{i-2}$, ..., $e_0$) is the sequence of edges representing the information flow from $u$ to $v$, and $l(e_*)$ and $l(u)$ represent the type of events and objects on the walk respectively. \textbf{Backward walk set} $W_i(v)$ of a given node $v$ is the set of all possible backward walks of length $i$ from $v$. 
\end{definition}

Each backward walk of length $i$ describes how node $v$ is impacted by the set of nodes $\{u\}$ with sequence of $i$ consecutive activities. For $i = 0$, the walk corresponds to the node itself, i.e. $(l(v))$. For instance, in Figure~\ref{fig:provsample}(a), length 2 backward walks for \texttt{registry1} are $(EDIT, CREATE, PROCESS)$ and $(EDIT, READ, FILE)$. Similarly, the walks of length 1 and 0 for \texttt{registry1} are $(PROCESS, EDIT)$ and $(REGISTRY)$ respectively. 

Given the set of backward walks $W_i(v)$ consisting of every length $i$ backward walks from node $v$, we group together labels at equal distances from $v$ in these walks. Formally, $\tau^j_i(v) = \{l(e_{i-j}) | \forall w \in W_i(v) \}$ for $1 \le j \le i$, and $\tau^0_i = \{l(u) | \forall w \in W_i(v)\}$, where each $w$ consists of a sequence of labels \{$l(e_0)$, $l(e_1)$, $l(e_2)$, ... $l(e_{i-1})$, $l(u)$\}. The labels $\tau^j_i$ for $0 \le j \le i$ are then stacked together to form \textbf{i-provenance label} , i.e., $\psi_i(v)$ = ($\tau^i_i$, $\tau^{i-1}_i$, ..., $\tau^1_i$, $\tau^0_i$). If no backward walk of length $i$ exists, then the empty set $\{\}$ is used to denote both $W_i(v)$ and i-provenance label $\psi_i(v)$. The process is repeated for each depth $i$ for $0 \le i \le h$. 
Let's look at the 0-, 1-, and 2-provenance labels of node \textit{registry1} in Figure~\ref{fig:provsample}(a),
\begin{itemize}
	\item For i = 0, $\psi_0(registry1) = (\{REGISTRY\})$, where $\tau^0_0 = \{REGISTRY\}$,
	\item For i = 1, $\tau^0_1(registry1) = \{PROCESS\}$, $\tau^1_1(registry1) = \{EDIT\}$, and $\psi_1(registry1) = (\{EDIT\}, \{PROCESS\})$
	\item For i = 2, $\tau^0_2(registry1) = \{PROCESS, FILE\}$, $\tau^1_2(registry1) = \{CREATE, READ\}$, $\tau^2_2(registry1) = \{EDIT\}$. Stacking all of them together, we get $\psi_2(registry1)$ = ($\{EDIT\}$, $\{CREATE, READ\}$, $\{PROCESS, FILE\}$).
\end{itemize}

For each provenance graph snapshot $G_t$, a histogram is constructed containing the frequency of different provenance labels for all nodes in the graph. The histogram keys are generated based on the unique $\psi_i(v)$ values for all $v \in V_t$ and $0 \leq i \leq h$, where $h$ is the maximum walk length. The histogram size of the provenance graph snapshots obtained in {\provtovec} is significantly smaller compared to the WL subtree kernel~\cite{shervashidze2011weisfeiler} and temporally sorted subtree kernel~\cite{han2020unicorn}.

In contrast to the multi-set approach used in the WL subtree kernel~\cite{shervashidze2011weisfeiler} and the temporally sorted multi-set approach in Unicorn~\cite{han2020unicorn}, the provenance kernel in {\provtovec} utilizes a set to aggregate labels from the neighborhood. This distinction is important because the multi-set approach has been deemed to provide better discrimination power necessary in many domains. However for provenance graph, since we use entity and event types as labels, these can generate spurious labels and weaken the generalization.

\begin{figure}[t]
	\begin{center}
		\includegraphics[width=0.45\textwidth ]{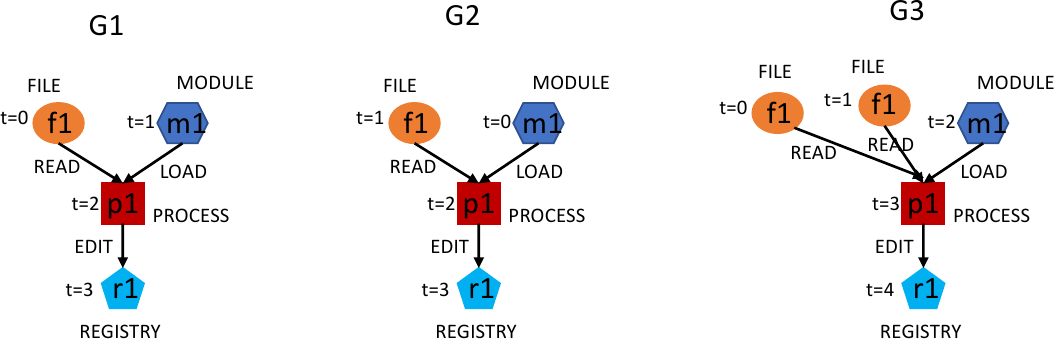}
		\caption{Three toy graphs representing a similar set of actions of a process reading from a file, loading module and editing a registry. Provenance graph kernel maps all of them to an identical histogram while existing kernels make distinction based on temporal order or repeated events.}
		\label{fig:toygraphs}
	\end{center}
	\vspace{-0.2in}
\end{figure}

For instance, take three graphs in Figure~\ref{fig:toygraphs} all of which represents a very similar set of actions, i.e., a process $p1$ reads from file(s), loads a module, and edits a registry item. In {\provtovec}, after mining length 1 backward walks, the same provenance label [$(LOAD,READ)$, $(FILE,MODULE)$] is generated for $p1$ in each of the graphs G1, G2, and G3. However, the WL-subtree kernel maps $p1$ in G3 to a different label [$(LOAD, MODULE)$, $(READ, FILE)$, $(READ, FILE)$] compared to $p1$ in G1 and G2, i.e., [$(LOAD, MODULE)$, $(READ, FILE)$]. Similarly, the Unicorn's kernel also considers the temporal order of $m1$ and $f1$, resulting in a different label for $p1$ in each graph. The ability of the provenance graph kernel to map similar behaviors to identical labels helps in better generalization of underlying behavior, reducing false positives in downstream tasks. This means that {\provtovec} can capture similarities between different instances of $p1$ across the graphs, while the other kernels may treat them as distinct. By providing consistent labels for similar behavior, the provenance graph kernel enhances the accuracy and effectiveness of subsequent analysis tasks.


\subsection{Incremental Provenance Graph Kernel}
Algorithm~\ref{alg:provkernel} presents a streaming approach for updating the provenance label histogram in real-time. It takes the newly inserted edges and iterates through them to obtain the provenance labels for newly inserted nodes and updated labels of impacted old nodes. First, it initializes the placeholders (lines 1 - 6) for provenance labels to hold $\psi_i(v)$ for all new nodes $v \in V_{new}$ and $0 \leq i \leq h$. In order to get $\psi_i(v)$, we need placeholder for $\tau^j_i(v)$ for $0 \leq j \leq i$. Once the initialization is done, we iterate through all inserted edges for $h$ times in order to obtain the provenance labels corresponding to the backward walks of length $0$ through $h$ (lines 7-17). Once the provenance labels are obtained, we update the label histogram to reflect the newly formed provenance labels(lines 14 - 17). 

In the graph snapshot of Figure~\ref{fig:provsample}(b), three new edges were inserted in the earlier snapshot, which creates three new nodes in the graph. Once the placeholders for $\psi_i$ and corresponding $\tau_i^j$ for each of these nodes are initialized, it obtains the provenance label of new nodes \textit{registry2}, \textit{process5.exe}, and \textit{IP2}, using the labels of their in-neighbors, i.e., (\textit{process2.exe}), (\textit{process3.exe}), and (\textit{process2.exe}) respectively. The new labels are then updated in the histogram. 

The runtime complexity of algorithm~\ref{alg:provkernel} is $\mathcal{O}(h^2|\Delta E| )$ for a given batch of edge insertions $\Delta E$. For the initial snapshot, the runtime complexity is $\mathcal{O}(h^2|E_0|)$, where $E_O$ represents the number of edges in the initial snapshot. The initialization phase (lines 1-6) can be completed in $\mathcal{O}(h^2|V_{new}|)$, where $V_{new}$ is the set of newly inserted nodes in the given snapshot, and the entire vertex set for the initial snapshot. After the initialization, the computation of provenance labels occurs in $h \times |\Delta E| \times h$ operations, as the process needs to update i-provenance labels for each of the inserted edge for $0 \leq i \leq h$. While the complexity is higher than $\mathcal{O}(h|\Delta E|)$ of WL subtree kernel~\cite{shervashidze2011weisfeiler} with h-hop neighborhood, it is important to note that the value of $h$ is typically very low (e.g., $\leq 4$). As a result, the overhead from the quadratic scaling is generally negligible in practice.

\begin{algorithm}[t]
	\caption{Incremental algorithm for computing provenance label histogram}
	\label{alg:provkernel}
	\KwData{Provenance graph snapshot $G_t$, current histogram $hist$, inserted edges $\Delta E$, new nodes $V_{new}$, max walk length $h$}
	\KwResult{An updated provenance label histogram $hist$}
	\Comment{Initialize the labels}
	\For{$v \in V_{new} $}{
		\For{$0 \leq i \leq h$}{
			$\psi_i(v) \leftarrow ()$\;
			\For{$0 \leq j \leq i$}{
				$\tau^j_i(v) \leftarrow \{ \}$;
			}
		}
		$\tau^0_0 \leftarrow \{l(v)\}$, $\psi_0 \leftarrow (\tau^0_0)$\;
	}
	\Comment {Iterate over inserted edges to infer other provenance labels}
	\For{$1 \leq i \leq h$}{
		\For{$e = (u, v) \in \Delta E$}{
			\If{$\psi_{i-1}(u)$ is empty}{
				skip the edge\;
			}
			$\tau^i_i(v)$.insert($l(e)$)\;
			\For{$0 \leq j \leq i-1$}{
				$\tau^j_i(v) \leftarrow \tau^j_i(v) \cup \tau^j_{i-1}(u)$\;
			}
			\Comment{Update the histogram, if there was a old label, we need to remove it}
			\If{$\psi_i(v)$ is not empty}{
				$hist[\psi_i(v)]--$\;
			}
			$\psi_i(v) = (\tau^i_i(v), \tau^{i-1}_i, ... , \tau^0_i)$\;
			$hist[\psi_i(v)]++$\;
		}
	}
\end{algorithm}

\subsection{Featurization of Histograms}
Most machine learning algorithms require a fixed-size input vector. The node label histograms from different snapshots have different number of bins, i.e., distinct node labels. We need to convert these variable size histograms to a fixed sized vectors. Let us assume histograms $H_0, H_1, ..., H_k$ are generated from graph snapshots $G_0, G_1, ..., G_k$. A label vocabulary $\Sigma$ is the set of all the distinct labels computed for all the nodes in all the graph snapshots, i.e., $\Sigma = \cup_{ i = 0}^k L_i$, where $L_i$ is the bins (labels) from $H_i$. 
%

%
In the streaming setting, where the label vocabulary is continuously expanding, we utilize a histosketch data structure to convert the variable-sized histogram $H_i$ into a fixed-size vector $S_i$ of size $K$. Histosketch employs consistent weighted hashing to transform the histogram into a compact sketch. By applying this technique, we can represent each snapshot with a fixed-size vector, regardless of the growing label vocabulary. To assess the similarity between two vectors $V_i \in \mathbf{R}^D$ and $V_j \in \mathbf{R}^D$, we can compute the distance between them using normalized min-max, which serves as a popular distance measure for non-negative vectors. Further details on histosketch can be found in Appendix A.

\begin{equation}
	D_{NMM}(V_i, V_j) = \frac{\sum_{l\in \Sigma}min(V_i[l], V_j[l])}{\sum_{l \in \Sigma}max(V_i[l], V_j[l])}
\end{equation}


\section{Evaluation}
\label{sec:p2vevaluation}
%

We utilized the x-stream edge-centric graph computing framework~\cite{roy2013xstream} to implement the graph kernels. This framework supports both in-memory and out-of-core graphs, enabling scalable computing on shared memory machines. In our implementation, node labels are stored on the vertices, and in each iteration of the graph kernel, the labels are scattered via edges and aggregated on the affected nodes to compute the set of newly formed labels from the streamed edges. This approach allows for efficient computation and maintenance of histograms and sketches in memory, while storing the provenance graph itself on disk. Other components of the {\provtovec} system, such as downstream task modeling and data parsing, were implemented using Python.

\noindent\textbf{Datasets: } We evaluated {\provtovec} in 3 different datasets:

\textbf{1. StreamSpot} dataset generated by~\cite{manzoor2016streamspot} contains information flow graphs derived from one attack and five benign scenarios. Each of the benign scenarios involves a normal task: watching Youtube, downloading files, browsing cnn.com, checking Gmail, and playing video games. The attack graphs are captured while a drive-by-download is triggered by visiting a malicious URL that exploits a flash vulnerability and gains root access to the visiting host. Each task is run 100 times on a \textbf{Linux machine} collecting a total of 600 graphs, where each graph encompasses all the system calls on the machine from boot up to shut down. In total, there are 5 different subject/object types and 29 different event types.

\textbf{2. SupplyChain attack scenarios} dataset~\cite{han2020unicorn} contains a whole system provenance including background activity captured by CamFlow (v0.5.0)~\cite{pasquier2017practical} while simulating two APT supply chain attacks SC-1 and SC-2 on a continuous integration (CI) platform. They follow a typical cyber kill chain with 7 non-exclusive phases, i.e., reconnaissance, weaponization, delivery, exploitation, installation, command and control(C\&C), and actions on objective~\cite{lockheedmartinckc}. In SC-1 GNU wget version 1.17 is exploited (CVE-2016-4971) using remote file upload when the victim requests a malicious URL to a compromised server. In SC-2, they exploited a vulnerability (CVE-2014-6271) from GNU Bash version 4.3, which allows remote attackers to execute arbitrary code via crafted trailing strings after function definitions in Bash scripts. Each scenario generates 125 graphs from the benign activity and 25 graphs from the attacker's activities. 

\textbf{3. Operational Transparent Cyber (OpTC)} data~\cite{dataset-optc} is collected over nine days at National Cyber Range in a simulated network with one thousand hosts, with half of the client machines turned off during data collection. Each host was running Windows 10 on VMware and was scripted to mimic daily user activities by performing common tasks such as creating, editing, and deleting word, powerpoint, excel, and text files; sending, receiving, and downloading files via emails; and browsing the internet. Three red-team APT exercises were performed, each on a separate day, where randomly chosen machines were targeted, compromised, and used to laterally move on to the other network clients. This dataset contains more than 17 billion events, from 500 hosts and 627 different users. Among these log events, there are 11 object types and 32 different event types. Most popular objects are FLOW (71.7\%),  FILE (12.4\%), PROCESS (8.6\%), MODULE (3.9\%), THREAD (3.0\%), and REGISTRY (0.3\%). The rest of the objects constitute less than 0.1\% of overall events. Only 0.3 million, approximately 0.0016\% of total events, are malicious~\cite{optc_utility}. 

\noindent\textbf{Graph Kernels: } Along with provenance graph kernel {\provtovec}, we implemented two other graph kernels from existing works. (1) Weisfeiler-Lehman Subtree kernel (\textbf{WLSubtree})~\cite{shervashidze2011weisfeiler} is implemented to include both edge labels and node labels in their aggregation. Using the edge and node label of each incoming neighbor of the given node $v$, a sorted multi-set of labels is built which is concatenated with the label of $v$. (2) The temporally ordered Weisfeiler-Lehman Subtree (\textbf{unicorn}) kernel~\cite{han2020unicorn} is implemented.

\noindent\textbf{Downstream Tasks:} We utilize the representation obtained from provenance graph kernel in three distinct downstream tasks:
\begin{itemize}
	\item \textbf{Graph classification} classifies the provenance graphs based on the underlying action being performed on the system. We use XGBoost classifier~\cite{xgboost} for graph classification. 
	\item \textbf{Novelty detection} using One-class support vector machine~\cite{oneclasssvm}. It is useful for detecting anomalous behavior in homogeneous system. 
	\item \textbf{Anomaly detection} using  K-Medoids Clustering. It uses partitioning around medoids (PAM) algorithm to minimize the distance between points labeled to be in a cluster and a point designated as the center of that cluster~\cite{kmedoids}. It is useful for detecting anomalous behavior in heterogeneous system, i.e., a system with multiple benign behavior profiles. 
\end{itemize} 
The average performance from 5-fold cross validation is reported in all of the prediction task reporting. The five fold split is only performed in benign graphs for the task of anomaly/novelty detection, i.e., four fifth of benign data are used to train the model.

%

 \begin{figure}[t]
	\begin{center}
		\includegraphics[width=0.45\textwidth ]{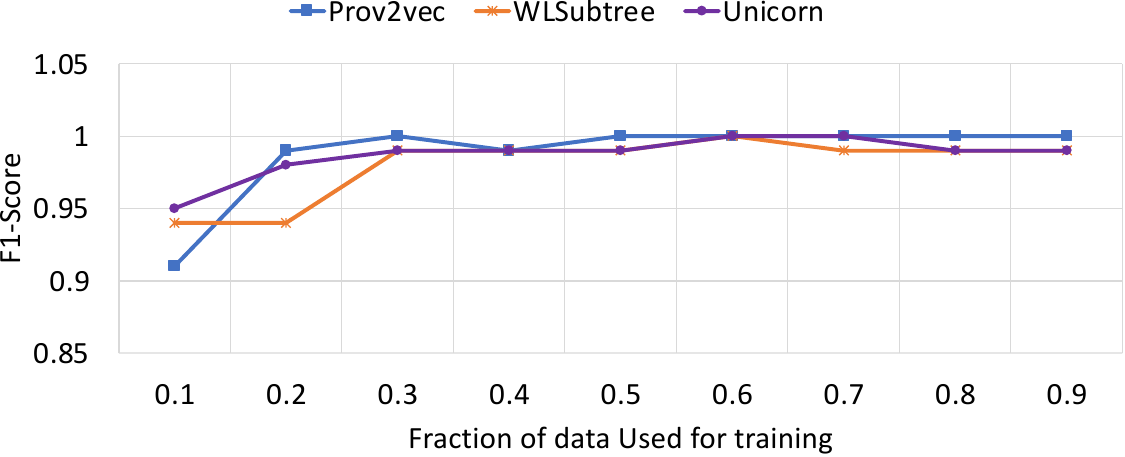}
		\vspace{-0.1in}
		\caption{The classification of graphs into 6 tasks(youtube, download, cnn, gmail, vgame, and attack) with varying amount of training data. }
		\label{fig:ss_classify}
	\end{center}
\end{figure}

\subsection{Graph Classification}
We obtained the static histograms on StreamSpot datasets, i.e., for each task and each run, one graph is built, and one histogram is constructed. We convert the histograms to sparse label frequency vectors, i.e., the feature vectors used here have sizes equal to the number of distinct node labels among all graphs, i.e., vocabulary size. 
We evaluate the ability of {\provtovec} to distinguish between different activities based on the provenance label histogram they generated. We use h = 3, i.e., the 3-hop neighborhood labels were collected for all of the different kernels. We use supervised learning by training XGB Classifier with a varying number of graphs and use the remaining graphs to test the classification performance. As depicted in Figure~\ref{fig:ss_classify}, all three kernel-based classifiers are able to reach peak classification performance in as little as around 20 graphs per task. This depicts the ability of the provenance kernel to identify similar tasks via a comparison of their provenance labels with a reasonable amount of data.

\subsection{Static Novelty Detection}
\begin{figure*}[t]
	\centering
	\begin{subfigure}[b]{0.28\textwidth}
		\centering
		\includegraphics[width=\textwidth]{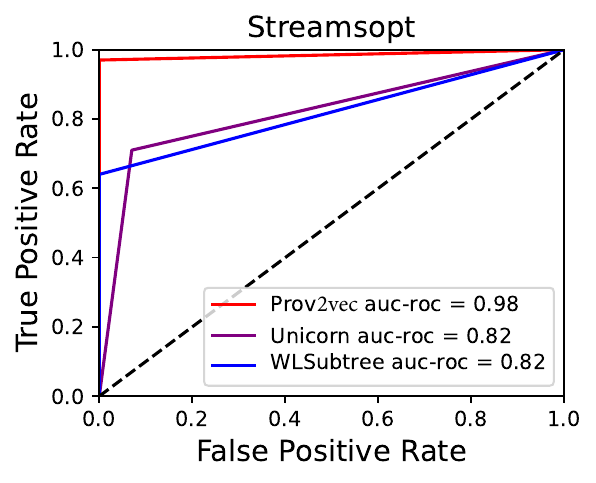}
		\label{fig:roc_prov}
	\end{subfigure}
	\hfill
	\begin{subfigure}[b]{0.28\textwidth}
		\centering
		\includegraphics[width=\textwidth]{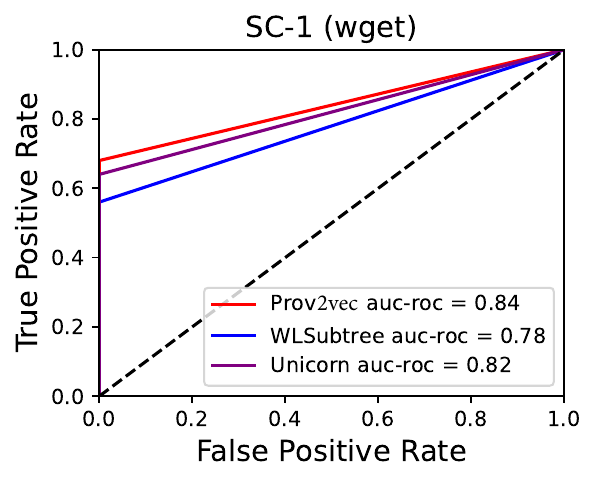}
		\label{fig:roc_wl}
	\end{subfigure}
	\hfill
	\begin{subfigure}[b]{0.28\textwidth}
		\centering
		\includegraphics[width=\textwidth]{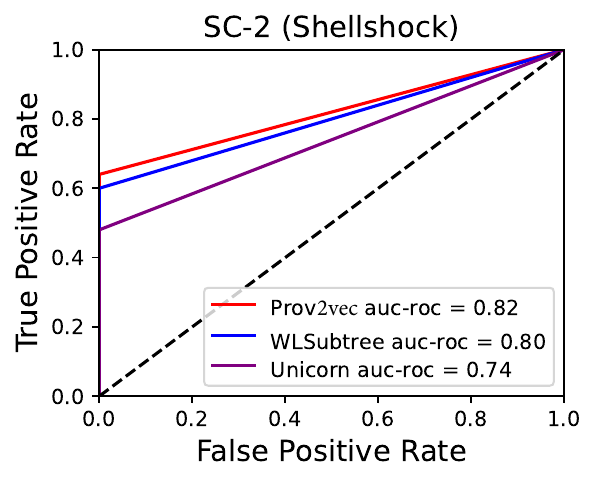}
		\label{fig:roc_unicorn}
	\end{subfigure}
	\vspace{-0.2in}
	\caption{ROC curve of one-class SVM based novelty detection for three different graph kernels on different datasets. The area under ROC curve for {\provtovec} kernel is consistently better than \textit{WLSubtree} and \textit{Unicorn} kernels.}
	\label{fig:roc_auc}
\end{figure*}

Using unsupervised learning, we predict the graphs that correspond to the attacks. We utilize 80\% of all benign tasks (400 graphs in StreamSpot and 100 graphs in SC-1 and SC-2) as normal behavior profiles and use them to train One-class SVM. The remaining 20\% of the benign activity graphs and all the graphs generated from the attack scenarios are used to test the anomaly detector, i.e., 200 graphs in StreamSpot and 50 graphs each in SC-1 and SC-2 respectively. Table~\ref{tbl:ss_anomaly} shows the performance for all three graph kernels and Figure~\ref{fig:roc_auc} shows the area under ROC curve for three kernel functions on the three datasets. Despite having a significantly smaller histogram size (Figure~\ref{fig:kernels}), the {\provtovec} outperforms both WLSubtree and time-ordered WL Subtree kernel from Unicorn~\cite{han2020unicorn}. The lower dimension of features helps the runtime of training and testing, while the better generalization of provenance using the concise histogram helps us to minimize the false positives, thereby improving the prediction ability of the anomaly detector.

\begin{table}[!ht]
	\centering
	\caption{The performance of one-class svm based anomaly detection on three different graph kernels(used h = 3 on each kernel). P, R, A, and F1 represents precision, recall, accuracy, and f1-score respectively. }
	\vspace{-0.1in}
	\begin{scriptsize}
		\begin{tabular}{|l|l|l|l|l|l|l|}
			\hline
			Dataset                     & Kernel    & P               & R            & A               & F1              & \begin{tabular}[c]{@{}l@{}}Runtime\\ (Sec)\end{tabular} \\ \hline
			\multirow{3}{*}{StreamSpot} & Prov2vec      & \textbf{0.9708} & \textbf{1.0} & \textbf{0.985}  & \textbf{0.9852} & \textbf{0.061}                                          \\ \cline{2-7} 
			& WLSubtree & 0.76            & 0.99         & 0.84            & 0.8609          & 1.281                                                   \\ \cline{2-7} 
			& Unicorn   & 0.7353          & 1.0          & 0.82            & 0.8475          & 3.034                                                   \\ \hline
			\multirow{3}{*}{SC-1}       & Prov2vec      & \textbf{0.7742} & \textbf{1.0} & \textbf{0.8571} & \textbf{0.8727} & \textbf{1.445}                                          \\ \cline{2-7} 
			& WLSubtree & 0.6857          & 1.0          & 0.7755          & 0.8136          & 5.281                                                   \\ \cline{2-7} 
			& Unicorn   & 0.7059          & 1.0          & 0.7959          & 0.8276          & 8.016                                                   \\ \hline
			\multirow{3}{*}{SC-2}       & Prov2vec      & \textbf{0.7353} & \textbf{1.0} & \textbf{0.82}   & \textbf{0.8475} & \textbf{1.751}                                          \\ \cline{2-7} 
			& WLSubtree & 0.7143          & 1.0          & 0.8             & 0.8333          & 10.687                                                  \\ \cline{2-7} 
			& Unicorn   & 0.6579          & 1.0          & 0.74            & 0.7937          & 14.539                                                  \\ \hline
		\end{tabular}
	\end{scriptsize}
	\label{tbl:ss_anomaly}
\end{table}


\subsection{Real-time Anomaly Detection }
The OpTC data provides a much better representation of real-world enterprise networks. The host logs for 500 different windows 10 hosts are collected over 9 days. During the first 6 days, only normal activities are performed on each host such as browsing the internet, playing video games, using Gmail, etc. Those 6 days are divided into 4 different boot-up to shut down sessions, i.e., (1) 17-18th, (2) 18-19th, (3) 19th, and (4) 20th - 23rd September 2019. We built different graphs for each host during each of these sessions, where the node label histogram is maintained incrementally and a snapshot is taken periodically. The series of histogram snapshots were then converted into fixed-sized sketch vectors of length 2048. All the sketches are then clustered using the k-medoid algorithm where an optimal number of clusters is determined by maximizing the silhouette coefficient~\cite{scikit-learn}. The trained k-medoid is then used for compromise detection during the evaluation period. 

The APT attack exercises were performed during the last 3 days, where one attack campaign is carried out each day. During the evaluation period, we create a provenance graph on each host every day and incrementally run graph kernels to compute node label histograms. The snapshots of histograms are taken periodically and are converted to sketch vectors. The resultant sketch vector is then tested against the k-medoids model trained during benign activity duration. If the sketch does not fit on any of the underlying clusters in the trained model, the snapshot is considered an anomaly. If a host in given evaluation day has at least one anomalous snapshot, we raise an alert indicating that the host has been compromised.

\begin{table}[!ht]
	\centering
	\caption{The anomaly detection results on 3 attack campaigns using k-medoids algorithm for h = 3 and sketch size = 2048. P, R, A, and F1 represents precision, recall, accuracy, and f1-score respectively.}
	\vspace{-0.1in}
	\begin{scriptsize}
		\begin{tabular}{|c|c|c|c|c|c|}
			\hline
			Attack                                                                                & Kernel    & P       & R          & A        & F1        \\ \hline
			\multirow{3}{*}{\begin{tabular}[c]{@{}c@{}}Day1- Powershell \\   Empire\end{tabular}} & Prov2vec      & \textbf{1.0000} & \textbf{0.1765} & \textbf{0.9720} & \textbf{0.3000} \\ \cline{2-6} 
			& WLSubtree & 0.6000          & 0.1765          & 0.9680          & 0.2727          \\ \cline{2-6} 
			& Unicorn   & 0.4000          & 0.1176          & 0.9640          & 0.1818          \\ \hline
			\multirow{3}{*}{Day2-Deathstar}                                                       & Prov2vec      & \textbf{1.0000} & \textbf{0.3333} & \textbf{0.9880} & \textbf{0.5000} \\ \cline{2-6} 
			& WLSubtree & 0.6667          & 0.2222          & 0.9840          & 0.3333          \\ \cline{2-6} 
			& Unicorn   & 0.3333          & 0.2222          & 0.9780          & 0.2667          \\ \hline
			\multirow{3}{*}{\begin{tabular}[c]{@{}c@{}}Day3-Malicious \\   Update\end{tabular}}   & Prov2vec      & \textbf{1.0000} & \textbf{1.0000} & \textbf{1.0000} & \textbf{1.0000} \\ \cline{2-6} 
			& WLSubtree & 0.2000          & 1.0000          & 0.9840          & 0.3333          \\ \cline{2-6} 
			& Unicorn   & 0.2857          & 1.0000          & 0.9900          & 0.4444          \\ \hline
		\end{tabular}
	\end{scriptsize}
	\label{tbl:optc_anomaly}
\end{table}

Table~\ref{tbl:optc_anomaly} shows the performance for detecting compromised hosts on each day of the attack. We used a time period of one hour between snapshots, neighborhood size of \textit{h=3} for graph kernels, and sketch the size of \textit{2048}. 
The precision represents the fraction of detected hosts that were actually compromised, while recall represents the fraction of compromised hosts that are detected. First, the precision of {\provtovec} kernel is much better than both \textit{WLSubtree} and \textit{unicorn} kernels. This is most likely down to the better generalization and a much more succinct histogram for {\provtovec} kernel compared to the other two techniques, which helps to provide a much better generalization of provenance for a given node. Notice that the recall is noticeably low for all of the kernels during day1 and day2. This is due to the fact that during these campaigns, there is hardly any activity on some of the compromised hosts where an attacker simply logs in after obtaining the credential from the domain controller. Below we discuss each of these attack campaigns in detail.

The attack campaign on day 1 uses PowerShell empire~\cite{empire}, where it manually connects to \textit{Sysclient201} as the user \textit{zleazer} and downloads malicious Powershell Empire stager.  It then uses privilege escalation methods to obtain elevated agents, Mimikatz to collect credentials, registry edits to establish persistence, and discovery techniques to gather system and network information. It then pivots to \textit{Sysclient402} using WMI invoke as an elevated agent, where it performs ping sweep of local network and pivots to \textit{Sysclient660}. Finally, it obtains domain controller information by using Powershell commands, pivots to \textit{DC1} (domain-controller 1), where it obtains the user hashes using lsa, and pivots to 14 different hosts. The detection process flags \textit{Sysclient201} and \textit{Sysclient660} as compromised with all three different kernels, while Unicorn kernel missed \textit{Sysclient402}. The remaining 14 hosts are missed as they do not have enough log data produced during the attacker's presence, and we could not flag the domain controller since there is no log collected for it.

The attack campaign of day2 was carried out using Deathstar, which starts with a phishing email containing malicious Powershell stagers to two users \textit{bantonio} and \textit{rsantilli}. On \textit{Sysclient501}, \textit{bantonio}  opens the malicious attachment. 
Once checked in the attacker runs a series of commands to list domain controllers, SID, and admins. It uses several UAC bypass techniques available in Powershell Empire such as \textit{eventvwr}, \textit{fodhelper}, \textit{wmi invoke}, and \textit{windir value modification} in order to escalate the privilege. It then starts reverse shell to the attacker, which downloads a netcat application with a different alias, compresses the content of \textit{Documents} folder into a file named \textit{export.zip} and copies it to news.com hosted at \textit{132.197.158.98}. The attacker pivots to \textit{Sysclient974} and explores files in the Documents folder. Similarly, it pivots to \textit{Sysclient005}, where it exfiltrates the data from the Downloads folder. The hosts \textit{Sysclient501}, \textit{Sysclient974}, and \textit{Sysclient005} are 3 out of 9 compromised hosts that are detected by all three kernels. 

On the day3, two hosts installed \textit{notepad.exe} susceptible to malicious upgrade, which when updated reaches out to the attacker's server hosted at \textit{53.192.68.50} and downloads a reverse tcp meterpreter payload that connects back to the attacker. Once connected, it runs discovery techniques to gather information on the local system, applications, domain controllers, and network shares. It then migrates to \textit{lsass} process, which uses \textit{Mimikatz} to collect clear-text passwords and hashes. Afterward, persistence is maintained by installing run keys and user `admin' is added to administrators and the RDP group. A similar approach was taken on both hosts \textit{Sysclient351} and \textit{Sysclient051}, where they leave large enough footprints for an anomaly detector to trigger the alert.

\begin{figure}[b]
	\begin{center}
		\vspace{-0.1in}
		\includegraphics[width=0.46\textwidth ]{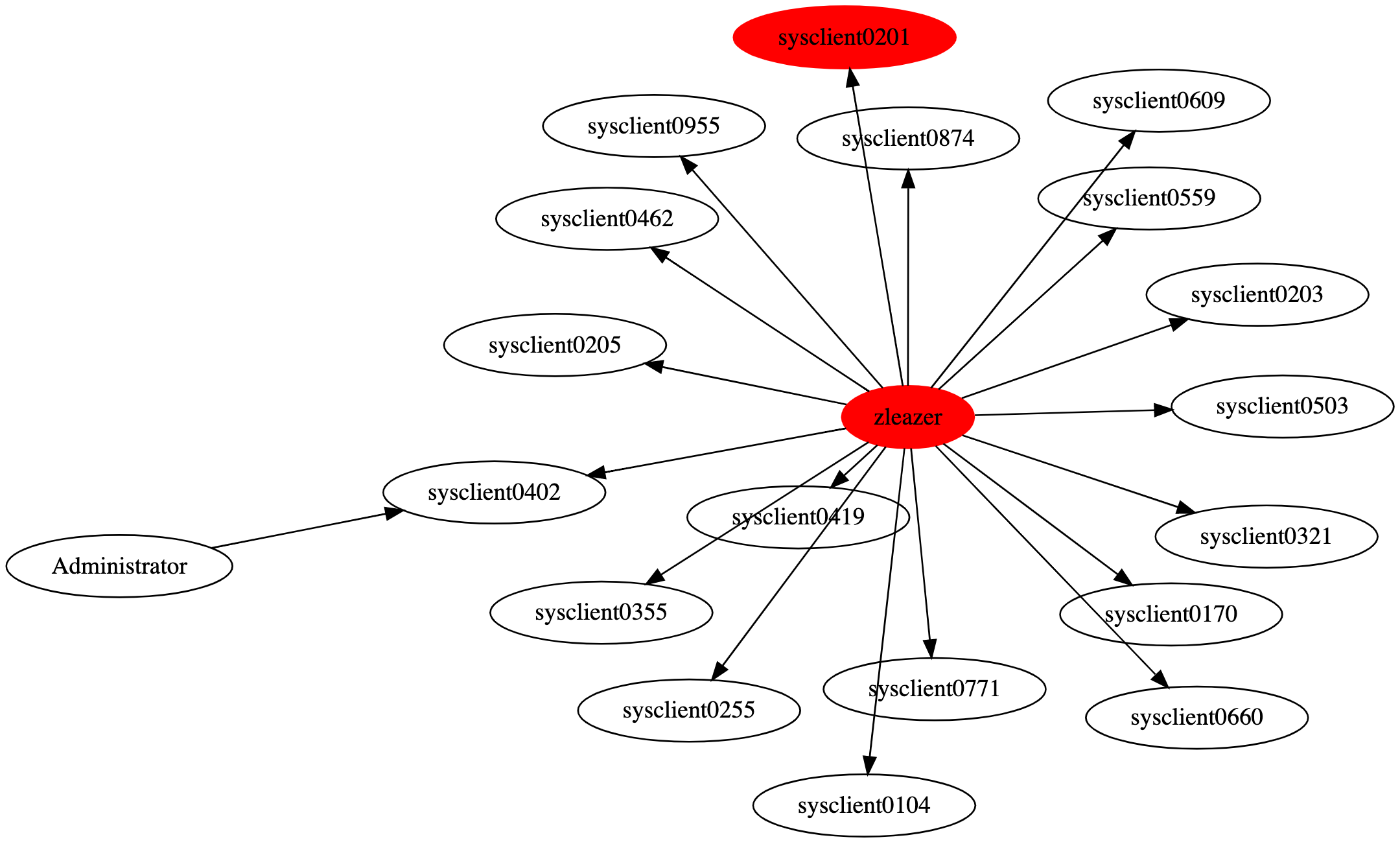}
		\caption{The movement of compromised user across network during attack campaign of day 1.}
		\label{fig:highlevelgraphs}
	\end{center}
\end{figure}

\begin{figure}[t]
	\begin{center}
		\includegraphics[width=0.44\textwidth ]{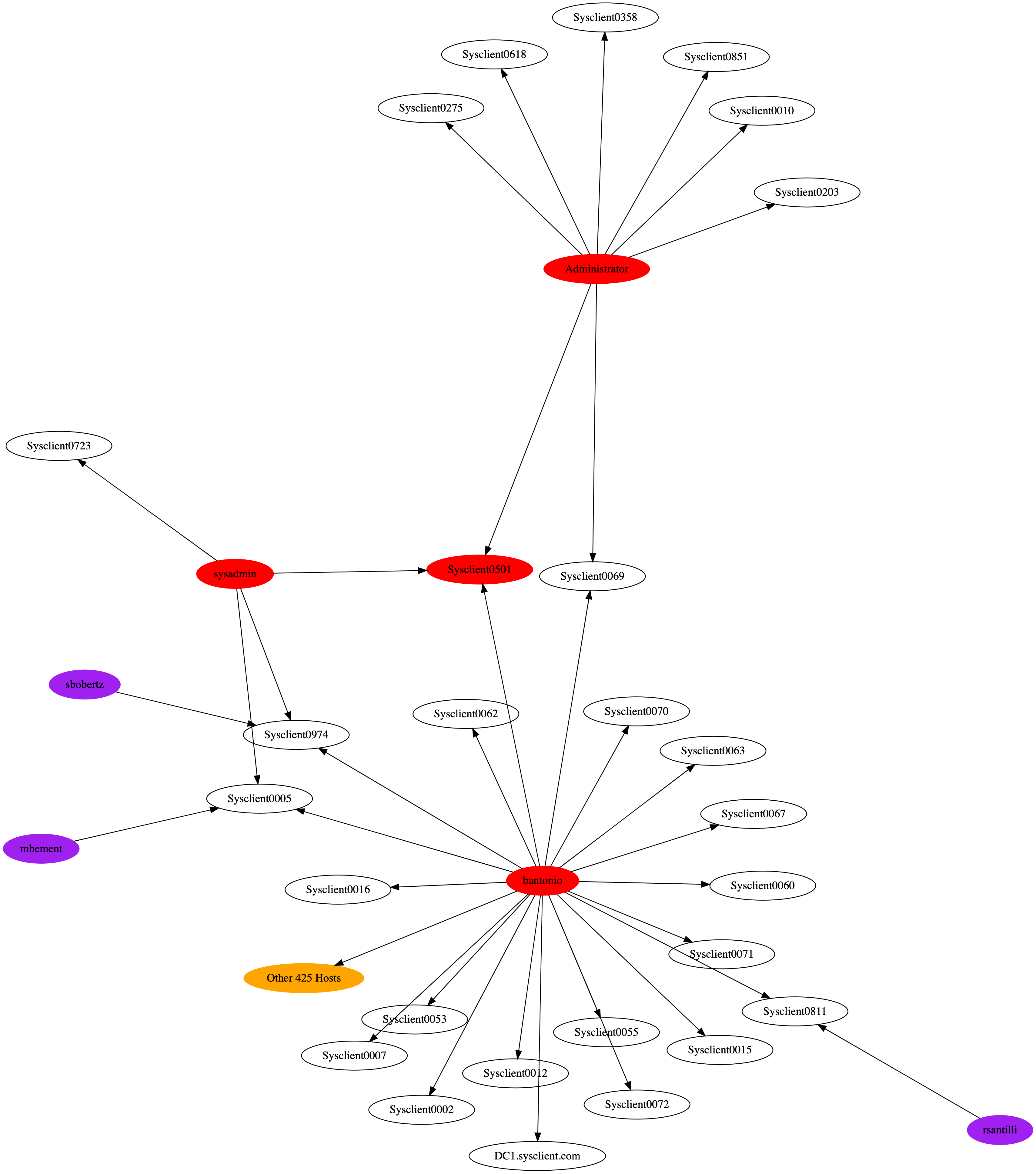}
		\vspace{-0.1in}
		\caption{The movement of compromised user across network during attack campaign of day 2.}
		\label{fig:highlevelgraphs1}
	\end{center}
\end{figure}

Afterward, we utilize a user-host interaction graph built using the user-session logs to flag potentially compromised hosts and users in order to quickly extract the impacted agents. The user-session logs in the OpTC data contains information such as user logins, logouts, and remote desktop protocol accesses and built a coarse-grained graph. When we detect a compromised host using the real-time anomaly detection on provenance graph snapshots, we extract the metadata from such anomalies, mainly the user, host, and the timestamp of the first anomaly. Following those agents and time information, we perform a temporal traversal on the user-host graph in order to obtain the potentially compromised hosts. Figure~\ref{fig:highlevelgraphs} and ~\ref{fig:highlevelgraphs1} shows the graphs containing the impacted hosts and users for the attack campaign of day1 and day2 respectively. With this temporal traversal, we were able to detect all the compromised hosts on day 1 except domain controller 1(DC1) as we did not have user-session logs for DC1. In addition it produced one false positive \textit{sysclient0203} which was not mentioned in ground truth. On day2, following this traversal obtained a bit large number of false positives as \textit{bantonio} logs into hundreds of hosts following the detection of an anomaly on \textit{Sysclient501}. However, the user with elevated privilege, i.e., \textit{Administrator} connects to all 9 hosts mentioned in the ground truth, which can be traced from the user-session logs. With this temporal traversal, we can detect the compromised hosts that were missed by anomaly detection as long as the anomaly detection finds at least one of the compromised hosts.

\begin{table}[b]
	\centering
	\caption{The evaluation of effect of different sized histosketches on the anomaly detection performance on StreamSpot data. P, R, A, and F1 represents precision, recall, accuracy, and f1-score respectively. K is sketch vector size.}
	\vspace{-0.1in}
	\begin{scriptsize}
		\begin{tabular}{|c|cccc|cccc|cccc|}
			\hline
			K & \multicolumn{4}{c|}{Prov2vec Kernel}                                                                                         & \multicolumn{4}{c|}{WLSubtree kernel}                                                                                     & \multicolumn{4}{c|}{Unicorn Kernel}                                                                                       \\ \hline
			& \multicolumn{1}{c|}{P}    & \multicolumn{1}{c|}{R}     & \multicolumn{1}{c|}{A}      & F1      & \multicolumn{1}{c|}{P}     & \multicolumn{1}{c|}{R}     & \multicolumn{1}{c|}{A}      & F1      & \multicolumn{1}{c|}{P}     & \multicolumn{1}{c|}{R}     & \multicolumn{1}{c|}{A}      & F1      \\ \hline
			32           & \multicolumn{1}{c|}{0.81}         & \multicolumn{1}{c|}{1}          & \multicolumn{1}{c|}{0.88}          & 0.89          & \multicolumn{1}{c|}{0.82}          & \multicolumn{1}{c|}{1}          & \multicolumn{1}{c|}{0.89}          & 0.9           & \multicolumn{1}{c|}{0.84}          & \multicolumn{1}{c|}{1}          & \multicolumn{1}{c|}{0.91}          & 0.91          \\ \hline
			64           & \multicolumn{1}{c|}{0.83}         & \multicolumn{1}{c|}{1}          & \multicolumn{1}{c|}{0.9}           & 0.9           & \multicolumn{1}{c|}{0.8}           & \multicolumn{1}{c|}{1}          & \multicolumn{1}{c|}{0.88}          & 0.89          & \multicolumn{1}{c|}{0.83}          & \multicolumn{1}{c|}{1}          & \multicolumn{1}{c|}{0.9}           & 0.9           \\ \hline
			128          & \multicolumn{1}{c|}{\textbf{0.9}} & \multicolumn{1}{c|}{\textbf{1}} & \multicolumn{1}{c|}{\textbf{0.94}} & \textbf{0.95} & \multicolumn{1}{c|}{0.83}          & \multicolumn{1}{c|}{1}          & \multicolumn{1}{c|}{0.9}           & 0.91          & \multicolumn{1}{c|}{0.76}          & \multicolumn{1}{c|}{1}          & \multicolumn{1}{c|}{0.85}          & 0.87          \\ \hline
			256          & \multicolumn{1}{c|}{0.9}          & \multicolumn{1}{c|}{1}          & \multicolumn{1}{c|}{0.94}          & 0.95          & \multicolumn{1}{c|}{0.88}          & \multicolumn{1}{c|}{1}          & \multicolumn{1}{c|}{0.93}          & 0.94          & \multicolumn{1}{c|}{0.85}          & \multicolumn{1}{c|}{1}          & \multicolumn{1}{c|}{0.91}          & 0.92          \\ \hline
			512          & \multicolumn{1}{c|}{0.89}         & \multicolumn{1}{c|}{1}          & \multicolumn{1}{c|}{0.94}          & 0.94          & \multicolumn{1}{c|}{\textbf{0.89}} & \multicolumn{1}{c|}{\textbf{1}} & \multicolumn{1}{c|}{\textbf{0.94}} & \textbf{0.94} & \multicolumn{1}{c|}{0.86}          & \multicolumn{1}{c|}{1}          & \multicolumn{1}{c|}{0.92}          & 0.92          \\ \hline
			1024         & \multicolumn{1}{c|}{}             & \multicolumn{1}{c|}{}           & \multicolumn{1}{c|}{}              &               & \multicolumn{1}{c|}{0.89}          & \multicolumn{1}{c|}{1}          & \multicolumn{1}{c|}{0.94}          & 0.94          & \multicolumn{1}{c|}{\textbf{0.89}} & \multicolumn{1}{c|}{\textbf{1}} & \multicolumn{1}{c|}{\textbf{0.94}} & \textbf{0.94} \\ \hline
			2048         & \multicolumn{1}{c|}{}             & \multicolumn{1}{c|}{}           & \multicolumn{1}{c|}{}              &               & \multicolumn{1}{c|}{0.89}          & \multicolumn{1}{c|}{1}          & \multicolumn{1}{c|}{0.94}          & 0.94          & \multicolumn{1}{c|}{0.89}          & \multicolumn{1}{c|}{1}          & \multicolumn{1}{c|}{0.94}          & 0.94          \\ \hline
		\end{tabular}
	\end{scriptsize}
	\label{tbl:sketch_impact}
	
\end{table}

\subsection{Effect of Sketch Size}
We evaluate the impact of using a fixed-size sketch vector in the performance of downstream tasks compared to the use of a sparse label histogram of size equal to the number of distinct labels among all graphs. 
We varied the size of the sketch from 32 to 2048, doubling each time to represent the node label histogram obtained by running all three kernels for $h = 3$. The histogram sketch obtained is thus used as the feature representation for the given graph. We trained the k-medoids clustering algorithm using 80\% of the graphs generated by benign activities. The remaining 100 benign graphs and 100 graphs generated during the attack are used for testing. During testing, each graph is tested against every cluster formed during training and flagged as an anomaly if it does not fit in any of the clusters. A graph is considered to fit in a cluster if its distance from the given clusters medoid is within $d$ standard deviation of the mean distance of all training samples in that cluster. In our experiments, we used $d = 2$, i.e., if a sample is farther than $mean+2 std$ away from all the medoids, it is considered an anomaly. The performance for varying sizes of sketches is shown in Table~\ref{tbl:sketch_impact} for anomaly detection on StreamSpot data.

The results in Table~\ref{tbl:sketch_impact} show that sketch size much smaller than the node label vocabulary size can match the performance for all kernels. The performance for {\provtovec} kernel saturates after a sketch size of 128. Similarly the performance for \textit{WLSubtree} and \textit{unicorn} kernels saturates at sketch size of  512 and 1024 respectively. The peak performance of \textbf{WLSubtree} and \textit{unicorn} kernels match that of their sparse histogram vector counterpart from Table~\ref{tbl:ss_anomaly}. However, the precision of {\provtovec} kernel is slightly amiss from its static counterpart. Nevertheless, sketching constantly changing and different-sized histograms with fixed-size feature sketches preserves the similarity between them and provide a viable option for comparing continuously changing provenance graphs.

\subsection{Effect of Neighborhood Size}
\begin{figure*}[t]
	\begin{center}
		\includegraphics[width=0.90\textwidth ]{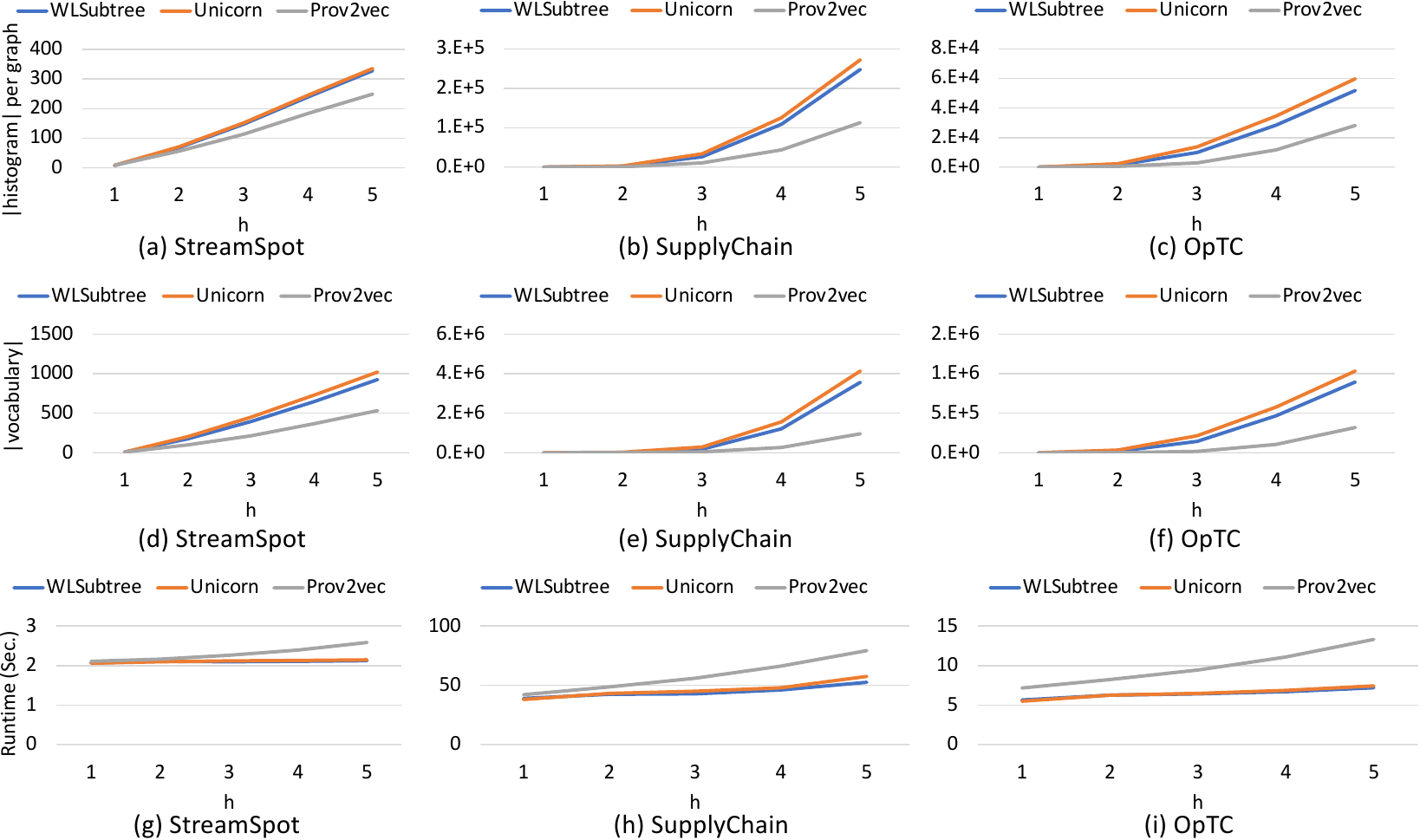}
		\vspace{-0.1in}
		\caption{The comparison of resource consumption for different kernels. The plots (a)-(c) shows the average size of histogram per graph, plots (d)-(f) shows the vocabulary size for different kernels, and plots (g)-(i) compares the runtime of different kernels for increasing neighborhood size.}
		\label{fig:kernels}
	\end{center}
\end{figure*}

\begin{figure}[b]
	\vspace{-0.1in}
	\begin{center}
		\includegraphics[width=0.45\textwidth ]{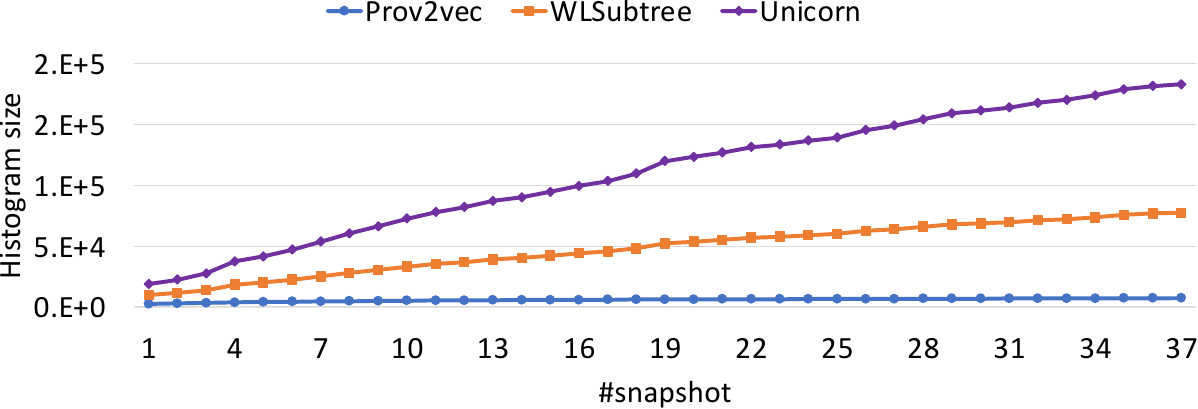}
		\vspace{-0.1in}
		\caption{The histogram size trend with each hourly snapshot on host 201 during 16-17Sep on OpTC data.}
		\label{fig:histpersnapshot}
	\end{center}
\end{figure}

We compared the resource consumption for using different kernels to compute the node label histograms in different datasets. We varied the value of h, i.e., the size of the neighborhood, and recorded the histogram size as well as the runtime for different graph kernels. As illustrated in Figure~\ref{fig:kernels}(a)-(f), the histogram for the 0-hop neighborhood is identical for all kernels, i.e., histograms built on node types. As the value of h increases, the difference between the sizes of histogram for \textit{unicorn} and \textit{WLSubtree} kernels compared to {\provtovec} kernel get larger. The comparison of histogram size growth over time for three kernels are shown in Figure~\ref{fig:histpersnapshot}. The number of labels and rate of arrival of unseen labels both are much smaller in provenance graph kernel. Despite this succinct representation, the performance on downstream task for {\provtovec} kernel is consistently better or comparable to the other two kernels as illustrated in the earlier subsection. 

The downside is increased runtime for provenance graph kernel as represented in Figure~\ref{fig:kernels}(g)-(i). Although the runtime for {\provtovec} kernel has quadratic growth with $h$, i.e., $h^2$ , compared to linear growth for \textit{WL Subtree} and \textit{unicorn} kernels, the optimal value of h is usually very small, thereby alleviating the impact of quadratic scaling.

Furthermore, we evaluated the impact of neighborhood size (h) based on the performance of corresponding histograms in downstream machine learning tasks. We used two SupplyChain datasets to evaluate the impact of neighborhood size on anomaly detection. We converted the histograms of corresponding snapshots to sketch vectors of size \textit{2048}. The performance for anomaly detection is shown in Table~\ref{tbl:himpact} for two attack scenarios SC-1 (wget) and SC2(shellshock). As expected, the performance for each kernel improves as we increase the neighborhood size, reach the peak for the value of h = 3 or 4, and start to decline afterward.

\section{Discussions and Limitations}
{\provtovec} makes certain assumptions and has limitations that should be considered.

First, it operates under the \textbf{closed-world assumption}, assuming that all benign behaviors have been observed during training~\cite{sommer2010outside}. However, in real enterprise networks, it is challenging to cover all possible benign cases. This may result in false alarms for previously unseen normal behaviors. To address this, system administrators can periodically update the model with new benign data. The incremental nature of {\provtovec} makes it easy for the model to update.


Second, {\provtovec} assumes an \textbf{integrity of training data} during a modeling period. It assumes that the newly observed normal behavior used for model updates is not corrupted by poisoning attacks~\cite{tang2020poisoning} or graph backdoors~\cite{xi2021backdoor}. The robustness of {\provtovec} against such attacks is an area for future study.


The \textbf{datasets used in the experiments are synthetic}, which limits the representation of real-world APT attacks. While efforts have been made to make the datasets realistic, they lack some characteristics of APT attacks in the wild. Testing {\provtovec} against actual enterprise systems or more realistic APT scenarios is a priority for future research.


\textbf{Granularity of data provenance: } Some attacks do not produce the attack pattern in the data provenance graphs. For example, malicious code in a file and thread-based attacks have the text information on the corresponding files and threads that are too fine granular to be recorded in the provenance graph. Like all provenance-based detection methods, {\provtovec} will fail to detect those attacks. Incorporating more host-based data into the threat detection process or improving the information capture process for finer-grained provenance graph generation can be the research directions to further investigate this limitation.

The explainability of anomalies is a challenge in black-box machine learning systems. {\provtovec} may struggle to provide detailed explanations for the detected anomalies. However, methods such as LIME and EDR systems can be used to explain individual predictions and understand the series of activities leading to an anomaly.

The provenance graph kernel \textbf{only supports discrete labels}, which limits its ability to capture continuous attributes. Including such attributes may require the use of deep learning techniques or graph kernels that support continuous attributes. Future work will explore whether these techniques can improve the performance of downstream prediction tasks.
Overall, while {\provtovec} has shown promising results, addressing these limitations will be crucial for its broader applicability and effectiveness in detecting sophisticated attacks.

\begin{table*}[t]
	\centering
	\caption{The evaluation of effect of different sized neighborhoods on the anomaly detection performance on SupplyChain data. P, R, A, and F1 represents precision, recall, accuracy, and f1-score respectively. }
	\vspace{-0.1in}
	\begin{scriptsize}
		\begin{tabular}{|c|c|cccc|cccc|cccc|}
			\hline
			&   & \multicolumn{4}{c|}{Prov2vec}                                                                                                            & \multicolumn{4}{c|}{WLSubtree}                                                                                                      & \multicolumn{4}{c|}{Unicorn}                                                                                                         \\ \hline
			\multirow{6}{*}{SC-1} & h & \multicolumn{1}{c|}{P}               & \multicolumn{1}{c|}{R}               & \multicolumn{1}{c|}{A}               & F1              & \multicolumn{1}{c|}{P}               & \multicolumn{1}{c|}{R}              & \multicolumn{1}{c|}{A}               & F1              & \multicolumn{1}{c|}{P}               & \multicolumn{1}{c|}{R}               & \multicolumn{1}{c|}{A}               & F1              \\ \cline{2-14} 
			& 1 & \multicolumn{1}{c|}{0.5333}          & \multicolumn{1}{c|}{0.3333}          & \multicolumn{1}{c|}{0.5306}          & 0.4103          & \multicolumn{1}{c|}{0.5333}          & \multicolumn{1}{c|}{0.3333}         & \multicolumn{1}{c|}{0.5306}          & 0.4103          & \multicolumn{1}{c|}{0.5333}          & \multicolumn{1}{c|}{0.3333}          & \multicolumn{1}{c|}{0.5306}          & 0.4103          \\ \cline{2-14} 
			& 2 & \multicolumn{1}{c|}{0.7778}          & \multicolumn{1}{c|}{0.875}           & \multicolumn{1}{c|}{0.8163}          & 0.8235          & \multicolumn{1}{c|}{0.7308}          & \multicolumn{1}{c|}{0.7917}         & \multicolumn{1}{c|}{0.7551}          & 0.76            & \multicolumn{1}{c|}{0.6667}          & \multicolumn{1}{c|}{0.8333}          & \multicolumn{1}{c|}{0.7143}          & 0.7407          \\ \cline{2-14} 
			& 3 & \multicolumn{1}{c|}{\textbf{0.8148}} & \multicolumn{1}{c|}{\textbf{0.9167}} & \multicolumn{1}{c|}{\textbf{0.8571}} & \textbf{0.8627} & \multicolumn{1}{c|}{0.7778}          & \multicolumn{1}{c|}{0.875}          & \multicolumn{1}{c|}{0.8163}          & 0.8235          & \multicolumn{1}{c|}{0.8148}          & \multicolumn{1}{c|}{0.9167}          & \multicolumn{1}{c|}{0.8571}          & 0.8627          \\ \cline{2-14} 
			& 4 & \multicolumn{1}{c|}{0.7333}          & \multicolumn{1}{c|}{0.9166}          & \multicolumn{1}{c|}{0.7959}          & 0.8148          & \multicolumn{1}{c|}{\textbf{0.84}}   & \multicolumn{1}{c|}{\textbf{0.875}} & \multicolumn{1}{c|}{\textbf{0.8571}} & \textbf{0.8571} & \multicolumn{1}{c|}{\textbf{0.8148}} & \multicolumn{1}{c|}{\textbf{0.9167}} & \multicolumn{1}{c|}{\textbf{0.8571}} & \textbf{0.8627} \\ \cline{2-14} 
			& 5 & \multicolumn{1}{c|}{0.75}            & \multicolumn{1}{c|}{0.875}           & \multicolumn{1}{c|}{0.7959}          & 0.8077          & \multicolumn{1}{c|}{0.8333}          & \multicolumn{1}{c|}{0.8333}         & \multicolumn{1}{c|}{0.8367}          & 0.8333          & \multicolumn{1}{c|}{0.7586}          & \multicolumn{1}{c|}{0.9167}          & \multicolumn{1}{c|}{0.8163}          & 0.8302          \\ \hline
			\multirow{6}{*}{SC-2} & h & \multicolumn{1}{c|}{P}               & \multicolumn{1}{c|}{R}               & \multicolumn{1}{c|}{A}               & F1              & \multicolumn{1}{c|}{P}               & \multicolumn{1}{c|}{R}              & \multicolumn{1}{c|}{A}               & F1              & \multicolumn{1}{c|}{P}               & \multicolumn{1}{c|}{R}               & \multicolumn{1}{c|}{A}               & F1              \\ \cline{2-14} 
			& 1 & \multicolumn{1}{c|}{0.5}             & \multicolumn{1}{c|}{0.04}            & \multicolumn{1}{c|}{0.5}             & 0.0741          & \multicolumn{1}{c|}{0.5}             & \multicolumn{1}{c|}{0.04}           & \multicolumn{1}{c|}{0.5}             & 0.0741          & \multicolumn{1}{c|}{0.5}             & \multicolumn{1}{c|}{0.04}            & \multicolumn{1}{c|}{0.5}             & 0.0741          \\ \cline{2-14} 
			& 2 & \multicolumn{1}{c|}{0.7222}          & \multicolumn{1}{c|}{0.52}            & \multicolumn{1}{c|}{0.66}            & 0.6047          & \multicolumn{1}{c|}{0.6}             & \multicolumn{1}{c|}{0.6}            & \multicolumn{1}{c|}{0.6}             & 0.6             & \multicolumn{1}{c|}{0.5862}          & \multicolumn{1}{c|}{0.68}            & \multicolumn{1}{c|}{0.6}             & 0.6296          \\ \cline{2-14} 
			& 3 & \multicolumn{1}{c|}{\textbf{0.7407}} & \multicolumn{1}{c|}{\textbf{0.8}}    & \multicolumn{1}{c|}{\textbf{0.76}}   & \textbf{0.7692} & \multicolumn{1}{c|}{\textbf{0.7143}} & \multicolumn{1}{c|}{\textbf{0.8}}   & \multicolumn{1}{c|}{\textbf{0.74}}   & \textbf{0.7547} & \multicolumn{1}{c|}{0.6552}          & \multicolumn{1}{c|}{0.76}            & \multicolumn{1}{c|}{0.68}            & 0.7037          \\ \cline{2-14} 
			& 4 & \multicolumn{1}{c|}{0.7727}          & \multicolumn{1}{c|}{0.68}            & \multicolumn{1}{c|}{0.74}            & 0.7234          & \multicolumn{1}{c|}{0.6333}          & \multicolumn{1}{c|}{0.76}           & \multicolumn{1}{c|}{0.66}            & 0.6909          & \multicolumn{1}{c|}{\textbf{0.75}}   & \multicolumn{1}{c|}{\textbf{0.72}}   & \multicolumn{1}{c|}{\textbf{0.74}}   & \textbf{0.7347} \\ \cline{2-14} 
			& 5 & \multicolumn{1}{c|}{0.75}            & \multicolumn{1}{c|}{0.48}            & \multicolumn{1}{c|}{0.66}            & 0.5853          & \multicolumn{1}{c|}{0.5909}          & \multicolumn{1}{c|}{0.52}           & \multicolumn{1}{c|}{0.58}            & 0.5532          & \multicolumn{1}{c|}{0.6552}          & \multicolumn{1}{c|}{0.76}            & \multicolumn{1}{c|}{0.68}            & 0.7037          \\ \hline
		\end{tabular}
	\end{scriptsize}
	\label{tbl:himpact}	
\end{table*}

\section{Related Works}
\textbf{Provenance graph} has been popular tool for threat hunting research in last few years. Several works have been proposed to improve the provenance data collection~\cite{pasquier2017practical,bates2015trustworthy,pohly2012hi},  redundancy elimination~\cite{lee2013loggc,ma2016protracer,xu2016high,hassan2018towards}, intrusion detection using provenance graphs~\cite{milajerdi2019holmes,milajerdi2019poirot,hossain2017sleuth,liu2019log2vec,han2020unicorn,gao2018aiql,gao2018saql,du2017deeplog,hassan2020tactical,hassan2020omegalog,bhattarai2022steinerlog}. We refer interested readers to the comprehensive survey on threat detection techniques using provenance graph~\cite{zipperle2022provsurvey}.

\textbf{Provenance query systems:} Traditional query systems are not optimized for provenance analysis. Several solutions~\cite{gao2018aiql,gao2018saql,shu2018threat,pasquier2018runtime} have been proposed to provide threat investigation specific abilities such as streaming queries, causality tracking, graph pattern matching, and anomaly analysis. These systems are implemented on top of mature stream processors or databases and take the provenance graphs specific data model and query engine.

\textbf{Provenance data reduction:} is very important for storage and computational efficiency. Causality preserving reduction~\cite{xu2016high} and subsequent dependence preserving reduction~\cite{hossain2018usenix} merge the events if they do not alter the causality or forward and backward reachability respectively. LogGC~\cite{lee2013loggc} proposes a provenance garbage collection, that finds the isolated "temporary" nodes and removes them. Since garbage collection and causality/dependency preserving reduction can remove correlation between alerts or alert themselves, we modified these reduction systems to preserve alerts.

\textbf{Threat detection with provenance graphs:} Sleuth~\cite{hossain2017sleuth} uses policy based rules to trigger alerts and uses \textbf{tag propagation} technique to store and transmit the system execution history. The \textbf{abnormal behavior detection} systems ~\cite{hassan2019nodoze,liu2018towards} learn host behavior from historical data or parallel systems and tries to find abnormal interaction between system entities. The \textbf{graph pattern matching and alignment} based works such as Holmes ~\cite{milajerdi2019holmes}, Poirot~\cite{milajerdi2019poirot}, Rapsheet~\cite{hassan2020tactical}, and SteinerLog~\cite{bhattarai2022steinerlog} uses indicator of attacks(IOAs) to generate suspicious events and chain them together using graph exploration techniques. They use those chain of alerts to detect the attacks as well as to reconstruct the individual steps taken by an attacker. However, a substantial amount of manual effort and domain expertise is required to come up with the relevant IOAs for matching. Eg., Poirot requires one to write a different query for each of the attack campaigns and find their alignment on a provenance graph. Holmes~\cite{milajerdi2019holmes}, Rapsheet~\cite{hassan2020tactical} and SteinerLog~\cite{bhattarai2022steinerlog} uses more fine-grained behavioral patterns representing different TTPs relevant to their system and follows the causal dependency in provenance graph to construct the attack campaigns. {\provtovec} follows the \textbf{graph embedding based systems} such as Unicorn~\cite{han2020unicorn} and Log2Vec~\cite{liu2019log2vec} closely, where it uses graph representation computation to embed the log entries and perform anomaly detection.  However, with more compact histogram and consequently better generalization, we are able to outperform Unicorn. 



Graph kernels are widely used for learning node and graph representations in machine learning tasks. These techniques iteratively accumulate and compress information from a node's neighborhood to derive a new node label. Various methods, such as random walks~\cite{kashima2003marginalized,vishwanathan2010graph,zhang2018retgk}, subtrees~\cite{ramon2003expressivity,shervashidze2011weisfeiler}, cyclic patterns~\cite{horvath2004cyclic}, shortest paths~\cite{borgwardt2005shortest}, and graphlets~\cite{prvzulj2007biological}, are employed to capture node neighborhoods. Recently, Graph Neural Networks (GNNs)~\cite{kipf2016semi,li2015gated,hamilton2017representation,velivckovic2017graph,xu2018powerful} have gained popularity for representation learning. GNNs utilize recursive aggregation to compute a node's representation vector by incorporating information from its neighborhood, with each iteration encompassing a larger one-hop neighborhood. Node representations are then aggregated to obtain the feature vector for the entire graph.

\textbf{Sequence-based learning} techniques, which involve converting log sequences into key vectors representing system events, have gained popularity in operational anomaly detection~\cite{vaarandi2015logcluster,hamooni2016logmine,messaoudi2018search}. Models based on recurrent neural networks (RNNs) or Transformers are then trained with these key sequences~\cite{du2017deeplog,du2019lifelong,debnath2018loglens}. During deployment, these models predict anomalous behavior by forecasting the next event based on the observed sequence. However, their effectiveness is limited as they mainly examine short system call sequences and struggle to capture long-term behavior, leaving them vulnerable to evasion techniques. To detect stealthy and slow Advanced Persistent Threat (APT) attacks, which require a broader context, graph-based techniques leveraging the causal relationships among events in provenance graphs offer more promising solutions.


\section{Conclusion}
We proposed a fully unsupervised technique in {\provtovec}, which was able to successfully learn the system host behaviors from their provenance graphs and identify the potentially malicious behaviors that differ from the normality. The provenance graph kernel proposed, while incurs a slight overhead in histogram computation compared to state-of-the-art graph kernels achieves an order magnitude smaller node label histogram sizes while improving the performance of downstream machine learning tasks at the same time. The result from {\provtovec} can be used as the first level of filtering for fine-grained alert correlation systems, where the anomalous hosts are further inspected to understand the context around underlying behavior.

\bibliographystyle{ACM-Reference-Format}
\bibliography{prov2vec}
\appendix
\section{Histosketch}
For a node label histogram $H$, where each label $l$ has a frequency value, $H[l] \geq 0$, consistent weighted sampling (CWS) sample $(l, a_l): 0 \leq a_l \leq H[l]$, which is both uniform and consistent. This CWS sample $(l, a_l)$ corresponds to the node label histogram bin $(l)$ and its scaled weight $(a_l)$ and is uniformly sampled from $\cup_l \{l\} \times [0, H[l]]$, meaning the probability of selecting label $l$ from $H$ is proportional to its label frequency in histogram, i.e., $H[l]$, and y is uniformly distributed on $[0, H[l]]$. The sample is also consistent, which means given two histograms $H_1$ and $H_2$, if $\forall l$, $H_1[l] \leq H_2[l]$, a sub-element $(l, a_l)$ is sampled from $H_1$ and satisfies $y \leq H_2[l]$, then $(l, a_l)$ will also be sampled from $H_2$~\cite{yang2017histosketch,li20150}.

For generating a consistent sample for a member of node label histogram, CWS uses three distributions using all labels from the histogram. For each $l \in H$, and $k= 1,2, ..., K$, CWS samples from $\gamma _{l,k} \approx \text{gamma(2,1)}$, $\beta _{l,k} \approx \text{uniform(0,1)}$, and $c_{l,k} \approx \text{gamma(2,1)}$. Once these distributions have been sampled, CWS generates a consistent sample for all labels in the histogram as follows.

\begin{equation}
	y_{l,k} = exp(log(H[l]) - \gamma _{l,k} \beta _{l,k})
\end{equation} 

\begin{equation}
	a_{l,k} = \frac{c_{l,k}}{y_{l,k} exp(\gamma _{l,k})}
\end{equation}

For each sketch element, it chooses a label i.e., $S_k = argmin_{l\in H}a_{l,k}$, and corresponding hash value $A_k = min_{l \in H} a_{l,k}$.   Equations 3 and 4 are used to hash a label $l$ in proportion with its weight. Given two sketches $S_i$ and $S_j$ constructed from two histograms $H_i$ and $H_j$ respectively, the collision probability in sketches is exactly the normalized min-max similarity between the histograms(or sparse label frequency vectors):

\begin{equation}
	Pr[S_i[k] = S_j[k]] = D_{NMM}(H_i, H_j)
\end{equation}
where $k = 1,2, ..., K$. The normalized min-max similarity between two histograms then can be approximated by computing the hamming distance between two sketches. In addition, the histosketch can support a real-time update where every single update in histogram can be reflected in sketch vector in $\mathcal{O}(K)$. For each distinct label $l$ in histogram $H$, it requires $\mathcal{O}(K)$ space to store the precomputed distributions $\gamma_{l,k}$, $\beta_{l,k}$, and $c_{l,k}$, where $k = 1,2,3, ... K$, thereby making the resultant space complexity $\mathcal{O}(K\times|H|)$. 

\subsection{Sparse label frequency} 
In a static setting, where all the graph snapshots are already available, we accumulate the unique labels from all the graph snapshots to build a label vocabulary $\Sigma$. Then we convert the histogram $H_i$ of each snapshot into a sparse vector $V_i$ of size equal to $|\Sigma|$, where the element at the $i^{th}$ index is the frequency of label $i$ in the histogram of the given snapshot. The vector elements corresponding to the labels not present in the snapshot are set to 0.

\begin{equation}
	V_i[j] = \begin{cases}
		H_i[l] & \text{if $l \in H_i$} \\
		0 & \text{otherwise}
		\end{cases}
\end{equation}

\balance
\end{document}